\pdfoutput=1

\documentclass[prl,twocolumn,showpacs,preprintnumbers,amsmath,amssymb,superscriptaddress,nofootinbib]{revtex4-1}
\usepackage{graphicx,array,multirow,color}

\renewcommand{\bar}{\overline}

\newcommand{\mi}{\raisebox{0.75pt}{\scalebox{0.75}{$\,-\,$}}}

\newcommand{\fwbox}[2]{\text{\makebox[#1][c]{$\hspace{-150pt}\displaystyle#2\hspace{-150pt}$}}}
\newcommand{\fwboxL}[2]{\text{\makebox[#1][l]{$#2$}}}
\newcommand{\fwboxR}[2]{\text{\makebox[#1][r]{$#2$}}}
\newcommand{\eq}[1]{\vspace{-0.pt}\begin{equation}\hspace{-100pt}#1\hspace{-100pt}\vspace{-0.pt}\end{equation}}

\renewcommand{\phi}{\varphi}
\DeclareMathOperator*{\Res}{\mathrm{Res}}

\newcommand{\cO}{\mathcal{O}}
\definecolor{dim}{rgb}{0.75,0.75,0.75}

\begin{document}
\title{Perturbation Theory at Eight Loops:\\Novel Structures and the Breakdown of Manifest Conformality\\[-20pt]}
\author{Jacob~L.~Bourjaily}
\affiliation{Niels Bohr International Academy \& Discovery Center, Niels Bohr Institute, University of Copenhagen, Denmark}
\author{Paul~Heslop}
\author{Vuong-Viet~Tran}
\affiliation{Department of Mathematical Sciences, Durham University, UK}
\preprint{DCPT-15/75}

\begin{abstract}
We use the soft-collinear bootstrap to construct the 8-loop integrand for the 4-point amplitude and 4-stress-tensor correlation function in planar maximally supersymmetric Yang-Mills theory. Both have a unique representation in terms of planar, conformal integrands grouped according to a hidden symmetry discovered for correlation functions. The answer we find exposes a fundamental tension between manifest locality and planarity with manifest conformality not seen at lower loops. For the first time, the integrand must include terms that are finite even on-shell and terms that are divergent even off-shell (so-called `pseudoconformal' integrals). We describe these novelties and their consequences in this letter, and we make the full correlator and amplitude available as part of this work's submission files to the {\tt arXiv}.
\end{abstract}
\maketitle

\section{introduction}\label{introduction_section}\vspace{-10pt}
Scattering amplitudes have been a rich source of theoretical data about the structure of quantum field theory, leading to the discovery of unanticipated simplicity and symmetry, and to the development of powerful new computational tools. This has especially been true in the case of amplitudes in planar, maximally supersymmetric (\mbox{$\mathcal{N}\!=\!4$}) Yang-Mills theory (`SYM'). Also of considerable interest, developed in parallel with amplitudes, has been the study of correlation functions of gauge invariant operators in the same theory. These are the fundamental objects appearing in the AdS/CFT correspondence.

The 4-point amplitude and correlation function in planar SYM have proven especially rich examples of perturbation theory. While the correlation function is strictly finite, the amplitude is of course (infrared) divergent. However, the loop {\it integrands} of both objects are completely well defined, and these are conjectured to be related by the duality discovered in \cite{Eden:2010zz,Alday:2010zy} and elaborated in \cite{Eden:2010ce,Eden:2011yp,Eden:2011ku,Adamo:2011dq}. To date, these integrands have been determined (using a variety of techniques) through 7 loops for both the amplitude \cite{Bern:1997nh,Bern:2005iz,Bern:2006ew,Bern:2007ct,Bourjaily:2011hi} and the correlator \cite{GonzalezRey:1998tk,Eden:1998hh,Eden:1999kh,Eden:2000mv,Bianchi:2000hn,Eden:2011we,Eden:2012tu,Ambrosio:2013pba,Drummond:2013nda}. 

In this letter, we extend the reach of this theoretical data to 8-loop order for both the amplitude and correlator using the so-called `soft-collinear bootstrap' \mbox{\cite{Bourjaily:2011hi}}, and we describe some of the surprising features that are found. It is worth emphasizing that without input from the correlator side of the duality, the soft-collinear bootstrap method applied to the amplitude alone would have failed beyond 7 loops. This is because, starting at 8 loops, there exist strictly finite conformal integrals---namely: \vspace{-6pt}\eq{\hspace{-5pt}\raisebox{-34.5pt}{\includegraphics[scale=1]{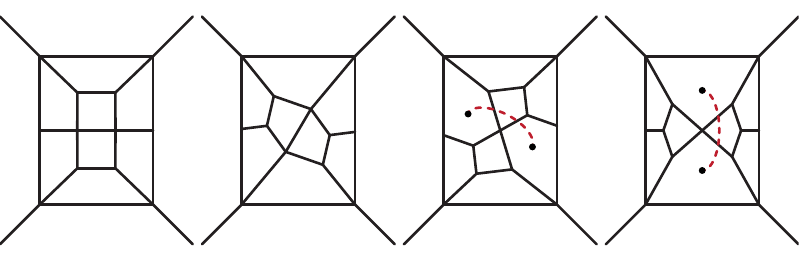}}\label{finite_graph_examples}\vspace{-6pt}}
These integrals are finite in the collinear limit, and so they do not contribute to the collinear divergence. Because of this, their contribution to the amplitude cannot be determined using the bootstrap without some additional input. This input is provided by the correlator side of the duality, in which every finite integral in (\ref{finite_graph_examples}) is related to one that does contribute to the collinear divergence, allowing its coefficient to be fixed. (We expect that this is the case for all finite terms at all loop-orders.) Using this hidden symmetry, we will find that all the integrals in (\ref{finite_graph_examples}) do in fact contribute to the 8-loop amplitude, with coefficients $\{-1,1/2,1/2,1\}$, respectively. 

The existence of strictly finite integrals such as those in (\ref{finite_graph_examples}) is one of the important novelties discovered at 8 loops. The other principle (and wholly unanticipated) novelty is the necessary contributions from so-called `pseudoconformal' (but not truly conformal) integrals such as:
\vspace{-6pt}\eq{\hspace{-5pt}\raisebox{-34.5pt}{\includegraphics[scale=1]{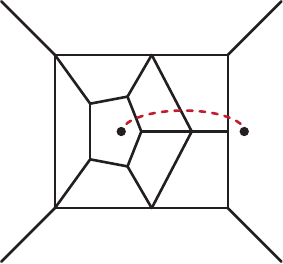}}\label{pseudo_conformal_example}\vspace{-6pt}}
These are conformal as {\it integrands}, but remain divergent even off-shell, spoiling the manifest finiteness (hence conformality) of the correlation function. Moreover, they remain divergent along the Higgs branch of the theory---preventing the use of the mass-regulator described in \cite{Alday:2009zm}.

We elaborate on both of these novelties and their consequences after first reviewing the amplitude/correlator duality and the methodology used to find the amplitude and correlation function. Complete expressions for both the amplitude and correlator are included as {\sc Mathematica} files in this work's submission to the {\tt arXiv}.

\vspace{-10pt}\section{The 4-point amplitude and correlator}\vspace{-10pt}
For 4 particles, the amplitude/correlator duality relates the scattering amplitude, $\mathcal{A}(x_i)$, expressed in dual-momentum coordinates (and divided by the tree) to the correlation function $\langle\cO(x_1)\bar{\cO}(x_2)\cO(x_3)\bar{\cO}(x_4)\rangle$. Here, the operator $\cO(x)$ is the trace over the gauge group of the square of one of the six scalars, $\cO(x)\!\equiv\!\mathrm{Tr}(\phi(x)^2)$, related via supersymmetry to the entire stress-tensor multiplet. The dual-momentum coordinates $x_i$ are related to ordinary momenta via \mbox{$p_i\!\equiv\!(x_{i+1}\mi x_{i})\!\equiv\!x_{i\,i+1}$}. For on-shell momenta, $p_i^2\!=\!x_{i\,i+1}^2\!=\!0$; but the correlation function is of course well defined for arbitrary $x_{i}\!\in\!\mathbb{R}^{3,1}$. With this, the amplitude and correlator are related via:
\vspace{-4pt}\eq{\hspace{-10pt}\lim_{x^2_{ii{+}1}\!\to0}\!\frac{\langle\cO(x_1)\bar{\cO}(x_2)\cO(x_3)\bar{\cO}(x_4)\rangle}{\langle\cO(x_1)\bar{\cO}(x_2)\cO(x_3)\bar{\cO}(x_4)\rangle_{\fwboxL{0pt}{\text{tree}}}}\,\,\,=\mathcal{A}(x_1,\ldots,x_4)^2.\label{eq:3}\vspace{-4pt}} 
As written here, both sides are of course divergent. However, the duality works at the level of the loop {\it integrand}, which is always well defined in a planar theory (upon symmetrization of the loop momenta). On both sides of the correspondence then, the $\ell$-loop integrand is some rational function on $(4{+}\ell)$ points in $x$-space. We will suggestively use $\{x_5,\ldots,x_{4+\ell}\}$ to denote the $x$-variables of the loop momenta.

Additionally, both sides of (\ref{eq:3}) are conformally invariant in $x$-space. For the correlator, this is the ordinary conformal invariance of \mbox{$\mathcal{N}\!=\!4$} SYM; but for the amplitude, this is the so-called `dual-conformal' invariance \cite{Drummond:2006rz}. Using dual-conformal symmetry, one can expand the amplitude into any complete basis of dual-conformal invariant (DCI) integrands, and fix their coefficients using unitarity, for example. Because the set of planar, dihedrally-symmetrized DCI integrands (with numerators involving products of `simple' Lorentz-invariants---of the form $x_{ij}^2$) forms a complete (and not over-complete) basis, the coefficient of any particular DCI integrand is well defined. That is, there is a unique representation of the amplitude in terms of DCI integrands, and we can meaningfully discuss `the' coefficient of an integrand such as that in (\ref{pseudo_conformal_example}).  

The expansion of the amplitude or correlator integrand into the basis of DCI terms turns out to be {vastly} simplified by the existence of a powerful, hidden symmetry (arising non-trivially from superconformal symmetry) that relates the internal and external variables \cite{Eden:2011we,Eden:2012tu}. The entire 4-point correlation function of any operator in the stress-tensor multiplet can be expressed in terms of a related function, denoted $f^{(\ell)}(x_1,\ldots,x_4;x_5,\ldots,x_{4+\ell})$ (see \cite{Eden:2000bk} for details). This hidden symmetry states that $f^{(\ell)}$ is a {\it fully-symmetric} function of the $x_i$---both external and internal! Before describing the precise connection between the amplitude and the function $f^{(\ell)}$, let us first discuss the space of functions into which $f^{(\ell)}$ can be expressed, and how they may be classified. 

Locality and conformality imply that $f^{(\ell)}$ must be a rational function involving factors $x_{ij}^2$ with weight $\mi4$ in all variables; and analyticity ensures that $f^{(\ell)}$ can have at most single poles in $x_{ij}^2$. Combining these with planarity and permutation invariance greatly restricts the space of possible functions into which $f^{(\ell)}$ may be expanded. We call these functions `$f$-graphs'. It is surprisingly easy to enumerate all possible $f$-graphs. Consider each factor $x_{ij}^2$ appearing in the denominator as the edge of a graph connecting $x_{i}\!\to\! x_j$. Then the space of possible denominators is simply the space of plane graphs involving $(4{+}\ell)$ vertices, each with valency $\geq\!4$ (due to the conformal weights). These can be rapidly enumerated (to high orders) using the program {\tt CaGe} \cite{CaGe}, for example. 

At 8 loops, for example, we find that there are $3,\!763$ 1-connected plane graphs (and counting distinct plane embeddings separately). For each of these possible $f$-graph denominators, we construct all (inequivalent) numerators involving the factors $x_{ij}^2$ that would result in a function with weight $\mi4$ in all variables. This is easy to do, and the result is a complete classification of $f$-graphs at $\ell$ loops. We have completed this classification exercise through 10 loops---statistics of which is summarized in \mbox{Table \ref{f_graph_statistics_table}}. 

\begin{table}[t]$\begin{array}{|@{$\,$}l@{$\,$}|@{$\,$}r@{$\,$}|@{$\,$}r@{$\,$}|@{$\,$}r@{$\,$}|}\multicolumn{1}{@{$\,$}c@{$\,$}}{\begin{array}{@{}l@{}}\text{$\ell\,$}\end{array}}&\multicolumn{1}{@{$\,$}c@{$\,$}}{\!\begin{array}{@{}c@{}}\text{\,\,\# of plane graphs\,\,}\end{array}\!}&\multicolumn{1}{@{$\,$}c@{$\,$}}{\!\begin{array}{@{}c@{}}\text{\,\,\# of $f$-graphs\,\,}\end{array}\,\,}&\multicolumn{1}{@{$\,$}c@{$\,$}}{\begin{array}{@{}c@{}}\text{\,\# of DCI integrands}\end{array}}\\\hline2&1&1&1\\\hline3&1&1&2\\\hline4&4&3&8\\\hline5&14&7&34\\\hline6&69&36&284\\\hline7&446&220&3,\!239\\\hline8&3,\!763&2,\!709&52,\!045\\\hline9&34,\!662&43,\!017&1,\!026,\!511\\\hline10&342,\!832&900,\!145&24,\!113,\!353\\\hline\end{array}$\vspace{-6pt}
\caption{Statistics of $f$-graphs and DCI integrands for $\ell\!\leq\!10$.\label{f_graph_statistics_table}}\vspace{-10pt}\end{table}

Let us briefly review the relationship between $f$-graphs and planar contributions to the amplitude. The precise connection between the amplitude and $f^{(\ell)}$ is:
\vspace{-6pt}\eq{\hspace{-10pt}\lim_{x^2_{ii{+}1}\!\to0}\big(\xi f^{(\ell)}\big)=2\mathcal{A}^{(\ell)}+2\mathcal{A}^{(\ell-1)}\mathcal{A}^{(1)}+\ldots,\label{eq:1}\vspace{-6pt}}
with $\xi\!\equiv\!(x_{12}^2x_{23}^2x_{34}^2x_{41}^2(x_{13}^2)^2(x_{24}^2)^2)$, and the RHS coming from expanding $\mathcal{A}(x_i)^2$ in powers of the coupling. 
Each term in the expansion of the RHS of (\ref{eq:1}) can be independently read-off from the $f$-graph, with the leading term being of primary importance, as it gives the
$\ell$-loop amplitude: choosing any square face of the graph describing the denominator of an
$f$-graph (possibly built from two triangles which share an edge) to
be labelled $\{x_1,\ldots,x_4\}$, multiplying by the factor $\xi$, and taking the lightlike limit, we obtain a planar DCI integrand
that should appear in the basis for the $\ell$-loop amplitude. Different choices of faces for the lightlike limit will result in very different looking graphs. For example:
\vspace{-8pt}\eq{\hspace{-5pt}\raisebox{-34.5pt}{\includegraphics[scale=1]{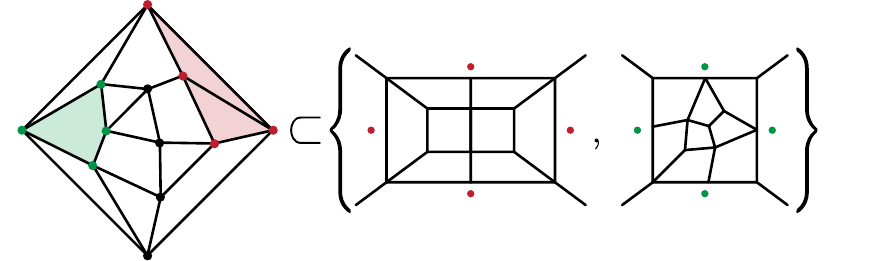}}\label{finite_and_divergent_pairing}\vspace{-8pt}}
Notice how these two apparently quite different planar DCI integrands (one of which is finite) are related as being different planar pieces of a single $f$-graph. Before moving on, it is worth mentioning that the extraction of planar DCI integrands from $f$-graphs is an incredibly efficient way to classify planar DCI integrands---the statistics of which have also been included in \mbox{Table \ref{f_graph_statistics_table}}.

\vspace{-10pt}\section{fixing coefficients}\label{sec:fixing-coefficients}\vspace{-10pt}
We used the so-called `soft-collinear bootstrap' to determine the coefficients of each $f$-graph in the expansion of the correlation function (via $f^{(\ell)}$)---equivalently, the coefficient of each planar DCI integrand (grouped into $f$-graph equivalence classes) in the expansion of the amplitude. Let us briefly review this approach (more thoroughly described in \mbox{ref.\ \cite{Bourjaily:2011hi}}). The key idea involved is the observation that the {\it logarithm} of the amplitude must be free of any soft-collinear divergence. By itself, this criterion seems quite weak; and yet, as has now been confirmed through 8 loops by direct computation, it turns out to be sufficient to uniquely determine the coefficient of every possible contribution to the amplitude or correlation function.

The soft-collinear region corresponds to the configuration where a loop variable, say $x_{5}$, becomes lightlike separated from any two (consecutive) external points, say $x_{1}$ and $x_2$. We can parameterize the divergence in this collinear region as the residue corresponding to $x_{15}^2\!\to\!0$ and $x_{25}^2\!\to\!0$. The precise premise of the soft-collinear bootstrap is the observation that this residue of the logarithm of the amplitude vanishes:
\vspace{-6pt}\eq{\Res_{\substack{\{x_{15}^2,x_{25}^2\}\to0}}\!\!\Big(\!\log\mathcal{A}\Big)=0.\label{bootstrap_criterion}\vspace{-6pt}}
Upon expanding the logarithm in powers of the coupling, this constraint should be satisfied at each order of perturbation theory. At $8$ loops, for example, the expansion of the logarithm is:
\vspace{-6pt}\eq{\hspace{-12pt}(\log\mathcal{A})^{(8)}\!=\!\mathcal{A}^{(8)}\!\!-\!\mathcal{A}^{(7)}\mathcal{A}^{(1)}\!\!-\!\mathcal{A}^{(6)}\mathcal{A}^{(2)}\!\!+\!\ldots-\!\frac{1}{8}\big(\mathcal{A}^{(1)}\big)^8\!\!.\label{8_loop_log}\vspace{-6pt}}
We can compute the collinear residue for every lower-loop contribution appearing in (\ref{8_loop_log}), and for every planar DCI integrand associated with each of the $2,\!709$ $f$-graphs. The constraint that the total residue be zero, (\ref{bootstrap_criterion}), then becomes a simple problem of linear algebra to find the coefficients of each $f$-graph.

We should emphasize that it is not at all clear why the bootstrap criterion (\ref{bootstrap_criterion})---which is a {\it necessary} property of the amplitude---should be {\it sufficient}. But the fact that it suffices follows from the observation (so far empirically true through 8 loops) that the space of collinear residues of all planar DCI integrands (gathered into equivalence classes according to $f$-graphs) are linearly independent. At least through 8 loops, the full amplitude/correlator is the unique combination of terms that satisfies the bootstrap criterion. A summary of the distribution of coefficients that are found is provided in \mbox{Table \ref{amplitude_coefficient_statistics_table}}.

Finally, let us note that in order for a DCI integrand to contribute to the collinear divergence, it must involve at least two propagators connecting a loop variable to external points. In ordinary momentum space, this corresponds to an external leg connected to the graph by a 3-point vertex. This explains why all the graphs in (\ref{finite_graph_examples}) are finite in the collinear limit: all external legs are connected to the graph via $4$-point vertices. This alone does not imply that the integrals are strictly finite---but the additional work required to see this is trivial. 

In fact, we expect that all $f$-graphs at all loop-orders contribute to the collinear divergence. Graphically, these divergences are associated with a triangular face in the graph of the denominator (connecting an internal point to two external points). We expect that every $f$-graph should have at least one triangular face adjacent to a square face. If so, it would imply that any strictly finite DCI integral will be in the same $f$-graph-equivalence-class as one with a collinear divergence. 

\begin{table}[t]$\hspace{-120pt}\begin{array}{|@{$\,$}l@{$\,$}|@{$$}r@{$$}|@{$$}r@{$$}|@{$$}r@{$$}|@{$$}r@{$$}|@{$$}r@{$$}|@{$$}r@{$\,$}|@{$$}r@{$$}|@{$$}r@{$$}|}
\multicolumn{1}{@{$\,$}c@{$\,$}}{\multirow{1}{*}{$$}}&\multicolumn{8}{@{$\,$}c}{\text{\# of $f$-graphs (DCI integrands) with coefficient:}}\\[1pt]\cline{2-9}%
\multicolumn{1}{@{$\,$}c@{$\,$}|@{}}{\ell}&\fwbox{45pt}{+1}&\fwbox{45pt}{-1}&\fwbox{23pt}{+2}&\fwbox{18pt}{-2}&\fwbox{31pt}{+1/2}&\fwbox{31pt}{-1/2}&\fwbox{23pt}{-3/2}&\fwbox{18pt}{-5}\\[-0.5pt]\cline{1-9}1&\fwboxL{45pt}{\,\fwboxR{13.5pt}{1}}\fwboxR{0pt}{(1)}\,&\fwboxL{45pt}{\,\fwboxR{13.5pt}{{\color{dim}0}}}\fwboxR{0pt}{{\color{dim}(0)}}\,&\fwboxL{23pt}{\,\fwboxR{4.5pt}{{\color{dim}0}}}\fwboxR{0pt}{{\color{dim}(0)}}\,&\fwboxL{18pt}{\,\fwboxR{4.5pt}{{\color{dim}0}}}\fwboxR{0pt}{{\color{dim}(0)}}\,&\fwboxL{31pt}{\,\fwboxR{8.5pt}{{\color{dim}0}}}\fwboxR{0pt}{{\color{dim}(0)}}&\fwboxL{31pt}{\,\fwboxR{8.5pt}{{\color{dim}0}}}\fwboxR{0pt}{{\color{dim}(0)}\!}&\fwboxL{23pt}{\,\fwboxR{4.5pt}{{\color{dim}0}}}\fwboxR{0pt}{{\color{dim}(0)}}\,&\fwboxL{18pt}{\,\fwboxR{4.5pt}{{\color{dim}0}}}\fwboxR{0pt}{{\color{dim}{(0)}}}\,\\\hline2&\fwboxL{45pt}{\,\fwboxR{13.5pt}{1}}\fwboxR{0pt}{(1)}\,&\fwboxL{45pt}{\,\fwboxR{13.5pt}{{\color{dim}0}}}\fwboxR{0pt}{{\color{dim}(0)}}\,&\fwboxL{23pt}{\,\fwboxR{4.5pt}{{\color{dim}0}}}\fwboxR{0pt}{{\color{dim}(0)}}\,&\fwboxL{18pt}{\,\fwboxR{4.5pt}{{\color{dim}0}}}\fwboxR{0pt}{{\color{dim}(0)}}\,&\fwboxL{31pt}{\,\fwboxR{8.5pt}{{\color{dim}0}}}\fwboxR{0pt}{{\color{dim}(0)}}&\fwboxL{31pt}{\,\fwboxR{8.5pt}{{\color{dim}0}}}\fwboxR{0pt}{{\color{dim}(0)}\!}&\fwboxL{23pt}{\,\fwboxR{4.5pt}{{\color{dim}0}}}\fwboxR{0pt}{{\color{dim}(0)}}\,&\fwboxL{18pt}{\,\fwboxR{4.5pt}{{\color{dim}0}}}\fwboxR{0pt}{{\color{dim}{(0)}}}\,\\\hline3&\fwboxL{45pt}{\,\fwboxR{13.5pt}{1}}\fwboxR{0pt}{(2)}\,&\fwboxL{45pt}{\,\fwboxR{13.5pt}{{\color{dim}0}}}\fwboxR{0pt}{{\color{dim}(0)}}\,&\fwboxL{23pt}{\,\fwboxR{4.5pt}{{\color{dim}0}}}\fwboxR{0pt}{{\color{dim}(0)}}\,&\fwboxL{18pt}{\,\fwboxR{4.5pt}{{\color{dim}0}}}\fwboxR{0pt}{{\color{dim}(0)}}\,&\fwboxL{31pt}{\,\fwboxR{8.5pt}{{\color{dim}0}}}\fwboxR{0pt}{{\color{dim}(0)}}&\fwboxL{31pt}{\,\fwboxR{8.5pt}{{\color{dim}0}}}\fwboxR{0pt}{{\color{dim}(0)}\!}&\fwboxL{23pt}{\,\fwboxR{4.5pt}{{\color{dim}0}}}\fwboxR{0pt}{{\color{dim}(0)}}\,&\fwboxL{18pt}{\,\fwboxR{4.5pt}{{\color{dim}0}}}\fwboxR{0pt}{{\color{dim}{(0)}}}\,\\\hline4&\fwboxL{45pt}{\,\fwboxR{13.5pt}{2}}\fwboxR{0pt}{(6)}\,&\fwboxL{45pt}{\,\fwboxR{13.5pt}{1}}\fwboxR{0pt}{(2)}\,&\fwboxL{23pt}{\,\fwboxR{4.5pt}{{\color{dim}0}}}\fwboxR{0pt}{{\color{dim}(0)}}\,&\fwboxL{18pt}{\,\fwboxR{4.5pt}{{\color{dim}0}}}\fwboxR{0pt}{{\color{dim}(0)}}\,&\fwboxL{31pt}{\,\fwboxR{8.5pt}{{\color{dim}0}}}\fwboxR{0pt}{{\color{dim}(0)}}&\fwboxL{31pt}{\,\fwboxR{8.5pt}{{\color{dim}0}}}\fwboxR{0pt}{{\color{dim}(0)}\!}&\fwboxL{23pt}{\,\fwboxR{4.5pt}{{\color{dim}0}}}\fwboxR{0pt}{{\color{dim}(0)}}\,&\fwboxL{18pt}{\,\fwboxR{4.5pt}{{\color{dim}0}}}\fwboxR{0pt}{{\color{dim}{(0)}}}\,\\\hline
5&\fwboxL{45pt}{\,\fwboxR{13.5pt}{5}}\fwboxR{0pt}{(23)}\,&\fwboxL{45pt}{\,\fwboxR{13.5pt}{2}}\fwboxR{0pt}{(11)}\,&\fwboxL{23pt}{\,\fwboxR{4.5pt}{{\color{dim}0}}}\fwboxR{0pt}{{\color{dim}(0)}}\,&\fwboxL{18pt}{\,\fwboxR{4.5pt}{{\color{dim}0}}}\fwboxR{0pt}{{\color{dim}(0)}}\,&\fwboxL{31pt}{\,\fwboxR{8.5pt}{{\color{dim}0}}}\fwboxR{0pt}{{\color{dim}(0)}}&\fwboxL{31pt}{\,\fwboxR{8.5pt}{{\color{dim}0}}}\fwboxR{0pt}{{\color{dim}(0)}\!}&\fwboxL{23pt}{\,\fwboxR{4.5pt}{{\color{dim}0}}}\fwboxR{0pt}{{\color{dim}(0)}}\,&\fwboxL{18pt}{\,\fwboxR{4.5pt}{{\color{dim}0}}}\fwboxR{0pt}{{\color{dim}{(0)}}}\,\\\hline6&\fwboxL{45pt}{\,\fwboxR{13.5pt}{15}}\fwboxR{0pt}{(129)}\,&\fwboxL{45pt}{\,\fwboxR{13.5pt}{10}}\fwboxR{0pt}{(99)}\,&\fwboxL{23pt}{\,\fwboxR{4.5pt}{1}}\fwboxR{0pt}{(1)}\,&\fwboxL{18pt}{\,\fwboxR{4.5pt}{{\color{dim}0}}}\fwboxR{0pt}{{\color{dim}(0)}}\,&\fwboxL{31pt}{\,\fwboxR{8.5pt}{{\color{dim}0}}}\fwboxR{0pt}{{\color{dim}(0)}}&\fwboxL{31pt}{\,\fwboxR{8.5pt}{{\color{dim}0}}}\fwboxR{0pt}{{\color{dim}(0)}\!}&\fwboxL{23pt}{\,\fwboxR{4.5pt}{{\color{dim}0}}}\fwboxR{0pt}{{\color{dim}(0)}}\,&\fwboxL{18pt}{\,\fwboxR{4.5pt}{{\color{dim}0}}}\fwboxR{0pt}{{\color{dim}{(0)}}}\,\\\hline7&\fwboxL{45pt}{\,\fwboxR{13.5pt}{70}}\fwboxR{0pt}{(962)}\,&\fwboxL{45pt}{\,\fwboxR{13.5pt}{56}}\fwboxR{0pt}{(904)}\,&\fwboxL{23pt}{\,\fwboxR{4.5pt}{1}}\fwboxR{0pt}{(7)}\,&\fwboxL{18pt}{\,\fwboxR{4.5pt}{{\color{dim}0}}}\fwboxR{0pt}{{\color{dim}(0)}}\,&\fwboxL{31pt}{\,\fwboxR{8.5pt}{{\color{dim}0}}}\fwboxR{0pt}{{\color{dim}(0)}}&\fwboxL{31pt}{\,\fwboxR{8.5pt}{{\color{dim}0}}}\fwboxR{0pt}{{\color{dim}(0)}\!}&\fwboxL{23pt}{\,\fwboxR{4.5pt}{{\color{dim}0}}}\fwboxR{0pt}{{\color{dim}(0)}}\,&\fwboxL{18pt}{\,\fwboxR{4.5pt}{{\color{dim}0}}}\fwboxR{0pt}{{\color{dim}{(0)}}}\,\\\hline8&\fwboxL{45pt}{\,\fwboxR{13.5pt}{472}}\fwboxR{0pt}{(9,\!047)}\,&\fwboxL{45pt}{\,\fwboxR{13.5pt}{434}}\fwboxR{0pt}{(9,\!018)}\,&\fwboxL{23pt}{\,\fwboxR{4.5pt}{8}}\fwboxR{0pt}{(67)}\,&\fwboxL{18pt}{\,\fwboxR{4.5pt}{1}}\fwboxR{0pt}{(7)}\,&\fwboxL{31pt}{\,\fwboxR{8.5pt}{78}}\fwboxR{0pt}{(923)}&\fwboxL{31pt}{\,\fwboxR{8.5pt}{63}}\fwboxR{0pt}{(869)\!}&\fwboxL{23pt}{\,\fwboxR{4.5pt}{3}}\fwboxR{0pt}{(17)}\,&\fwboxL{18pt}{\,\fwboxR{4.5pt}{1}}\fwboxR{0pt}{(1)}\,\\\hline\end{array}\hspace{-120pt}$\vspace{-6pt}
\caption{Amplitude/correlator coefficients through $8$-loops.\label{amplitude_coefficient_statistics_table}}\vspace{-10pt}\end{table}

\vspace{-10pt}\section{Results and Discussion}\label{sec:results-discussion}\vspace{-10pt}
The representation of the 8-loop integrand that has been found for the correlation function and amplitude includes two key novelties: the appearance of integrals that are finite even on-shell, and integrals that remain divergent even off-shell. Neither of these contributions were present at lower loop-orders, and they signal a fundamental tension between the properties and symmetries that the amplitude and correlation function are known to possess, and the ability to make these features manifest term-by-term. Let us briefly review each of these novelties in turn. 

Perhaps the most surprising new feature is the contribution from pseudoconformal integrals, such as that shown in (\ref{pseudo_conformal_example}). While conformal at the integrand-level, these terms obscure the ultimate conformality of the correlation function due to the presence of divergences that must be regularized. We have checked that the divergences of the pseudoconformal contributions cancel in combination; but it is quite surprising that the ultimate finiteness of the correlation function cannot be made manifest term-by-term (without breaking manifest locality or planarity).  

Although there do exist pseudoconformal integrals at lower loop-orders (starting at $\ell\!=\!5$), they do not contribute to the amplitude. Indeed, it has even been conjectured that they never do contribute---but we have seen this conjecture to fail 8 loops. Let us briefly review the structure of these pseudoconformal divergences, and how the amplitude/correlator duality provides an alternative explanation for their absence at lower loop-orders, while still allowing for their appearance at 8 loops. 

Divergences in a pseudoconformal integral can arise when some number, $k$, of the loop variables $x_{i\in I}$ approach another variable $x_j$ (either internal or external). Parameterizing the difference between each $x_{i\in I}$ and $x_j$ to be $\mathcal{O}(\epsilon)$, there will be a pole of order $\epsilon^{2E}$ in the denominator, where $E$ is the number of edges connecting the $(k{+}1)$ vertices in the set $I\!\cup\!\{j\}$ (minus the number of edges connecting vertices in this set appearing in the numerator). Going to polar coordinates for the $k$ integration variables $x_{i\in I}$ gives us an integrand proportional to $d\epsilon\,\epsilon^{4k-1}/\epsilon^{2E}$, which is divergent whenever $E\!\geq\!2k$. 

\begin{figure}[t]\caption{Subgraphs leading to pseudoconformal divergences.\label{pseudoconformal_divergences}}\vspace{-16pt}$$\hspace{4pt}\begin{array}{@{}c@{}}\raisebox{-0pt}{\includegraphics[scale=1]{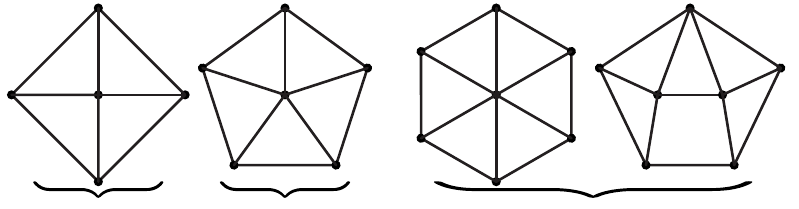}}\\[-7.5pt]\fwboxL{174pt}{\fwbox{0pt}{k=4}\hspace{53.75pt}\fwbox{0pt}{k=5}\hspace{101.15pt}\fwboxR{0pt}{k=6}}\end{array}\vspace{-22pt}$$\end{figure}

It is easy to classify the subgraphs that can lead to such a divergence. For $k\!=\!4$ through $k\!=\!6$, these are drawn in \mbox{Figure \ref{pseudoconformal_divergences}}. Importantly, in order for such a subgraph to signal a divergence, the numerator cannot involve any factors connecting the vertices of the subgraph to itself. (Such a numerator would remove the divergence by the power counting discussed above.)

\begin{figure}[t]\caption{$f$-graph manifestation of the rung-rule.\label{rung_rule}}\vspace{-16pt}$$\hspace{4pt}\begin{array}{@{}c@{}}\raisebox{-0pt}{\includegraphics[scale=1]{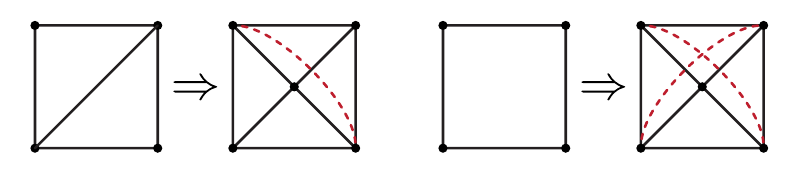}}\end{array}\vspace{-30pt}$$\vspace{0pt}\end{figure}

The simplest possible pseudoconformal divergence (first appearing at $5$ loops) is for $k\!=\!4$. Notice that this subgraph is very similar to the one relevant to the so-called `rung-rule', reviewed (in graphical form) in \mbox{Figure \ref{rung_rule}}. Interestingly, there is a strong reason why any $f$-graph containing a $k\!=\!4$ divergent subgraph cannot contribute to the correlator. Specifically, it would generate a term where the four points on the edge of the subgraph are taken lightlike, giving a contribution to $\mathcal{A}^{(1)}\mathcal{A}^{(\ell-1)}$ (since there is one point on the inside, and $(\ell{-}1)$ points outside the $4$-cycle). But such a term cannot be present at $(\ell{-}1)$-loops, since the corresponding $(\ell{-}1)$-loop $f$-graph would be non-planar, leading to a contribution. Another way to say this is that the term does not arise from the rung-rule on a (planar) lower-loop $f$-graph, and therefore cannot contribute to $f^{(\ell)}$. This logic provides a robust explanation of the absence of pseudoconformal contributions below 8 loops. 

The pseudoconformal contributions that appear at 8 loops all involve divergences arising from subgraphs with $k\!>\!4$. Such divergences cannot be excluded by the arguments from the amplitude/correlator duality given above. Indeed, we find that there are precisely 60 $f$-graphs that contribute at 8 loops (all with $k\!=\!5$ divergent subgraphs); and going to the lightlike limit, these 60 $f$-graphs contain a total of 560, planar DCI integrands that are individually divergent off-shell. 

A further intriguing feature of the 8-loop result is the appearance of new coefficients. Up to 7 loops only the coefficients $\pm 1$ and $2$ appeared whereas at 8 loops we see new integer coefficients ${-}2, {-}5$ as well as new half-integer coefficients: $\{{-}1/2,1/2,{-}3/2\}$. There is a single $f$-graph with coefficient ${-}5$, and it is also the first example of a graph with a hexagonal face. Indeed this follows a pattern: the  introduction of new coefficients has always accompanied new polygonal faces for the $f$-graphs. The first appearance of the coefficient $-1$ (at 4 loops) came with the first graph with a square face, and  the first appearance of $2$ (at 6 loops) accompanied the first graph with a
pentagonal face. The half integer coefficients which also appear for the first time at 8 loops are not so clearly distinguished. However, we can say that none of them contains a subgraph with a pentagon containing two vertices. The coefficient of such a graph is determined by the duality between the correlator and the 5-point amplitude. It seems that such graphs have coefficients inherited from lower loop graphs up to a sign. The fact that all the new coefficients can not be determined by a 5-point duality gives further evidence for this.

The other striking novelty of 8 loops is the contributions of finite integrals. These are unusual for a number of reasons, including the appearance of elliptic cuts (ultimately absent from the complete amplitude). To see this, consider the first graph appearing in equation (\ref{finite_graph_examples}); this graph contains a double-box with six massive (off-shell) legs. As pointed out in~\cite{CaronHuot:2012ab}, this implies that the diagram is not a expressible in terms of generalized polylogarithms. It is interesting that this structure, important for 10-point amplitudes at 2 loops \cite{Bourjaily:2015jna}, has some manifestation for 4 particles at 8 loops---illustrating the connections between many loops and many legs. 

Let us conclude by noting that there exists an  alternative approach to determining the correlation function. This involves the coincident limit~\cite{Eden:2012tu} (which can be rephrased as a simple graphical procedure on the $f$-graphs) in conjunction with information which can be obtained from the amplitude/correlator duality (which yields the rung-rule as well as a 5-point generalization suggested in~\cite{Ambrosio:2013pba}). Initial investigations indicate that these ideas are sufficient to completely fix the result to 8 loops; but we leave a more thorough discussion of these ideas to future work.

\newpage
\acknowledgments
The authors gratefully acknowledge the insights shared through conversations with Simon Caron-Huot, Burkhard Eden, Gregory Korchemsky,  Emery Sokatchev, Marcus Spradlin, and Jaroslav Trnka, and for custom computer code provided by Simon Caron-Huot for fast numerical evaluation. This work was supported in part by the Harvard Society of Fellows, a grant by the Harvard Milton Fund, and MOBILEX grant from the Danish Council for Independent Research (J.B.); by an STFC studentship (V.V.T.); and by a STFC Consolidated Grant ST/L000407/1 and the Marie Curie network GATIS (gatis.desy.eu) of the European Union's Seventh Framework Programme FP7/2007-2013/ under REA Grant Agreement No.\ 317089 (P.H.).


\begin{thebibliography}{10}

\bibitem{Eden:2010zz}
B.~Eden, G.~P. Korchemsky, and E.~Sokatchev,
\newblock JHEP {\bf 1112}, 002 (2011), arXiv:1007.3246.

\bibitem{Alday:2010zy}
L.~F. Alday, B.~Eden, G.~P. Korchemsky, J.~Maldacena, and E.~Sokatchev,
\newblock JHEP {\bf 1109}, 123 (2011), arXiv:1007.3243.

\bibitem{Eden:2010ce}
B.~Eden, G.~P. Korchemsky, and E.~Sokatchev,
\newblock Phys. Lett. {\bf B709}, 247 (2012), arXiv:1009.2488.

\bibitem{Eden:2011yp}
B.~Eden, P.~Heslop, G.~P. Korchemsky, and E.~Sokatchev,
\newblock Nucl. Phys. {\bf B869}, 329 (2013), arXiv:1103.3714.

\bibitem{Eden:2011ku}
B.~Eden, P.~Heslop, G.~P. Korchemsky, and E.~Sokatchev,
\newblock Nucl. Phys. {\bf B869}, 378 (2013), arXiv:1103.4353.

\bibitem{Adamo:2011dq}
T.~Adamo, M.~Bullimore, L.~Mason, and D.~Skinner,
\newblock JHEP {\bf 1108}, 076 (2011), arXiv:1103.4119.

\bibitem{Bern:1997nh}
Z.~Bern, J.~Rozowsky, and B.~Yan,
\newblock Phys. Lett. {\bf B401}, 273 (1997), arXiv:hep-ph/9702424.

\bibitem{Bern:2005iz}
Z.~Bern, L.~J. Dixon, and V.~A. Smirnov,
\newblock Phys. Rev. {\bf D72}, 085001 (2005), arXiv:hep-th/0505205.

\bibitem{Bern:2006ew}
Z.~Bern, M.~Czakon, L.~J. Dixon, D.~A. Kosower, and V.~A. Smirnov,
\newblock Phys. Rev. {\bf D75}, 085010 (2007), arXiv:hep-th/0610248.

\bibitem{Bern:2007ct}
Z.~Bern, J.~Carrasco, H.~Johansson, and D.~Kosower,
\newblock Phys. Rev. {\bf D76}, 125020 (2007), arXiv:0705.1864.

\bibitem{Bourjaily:2011hi}
J.~L. Bourjaily, A.~DiRe, A.~Shaikh, M.~Spradlin, and A.~Volovich,
\newblock JHEP {\bf 1203}, 032 (2012), arXiv:1112.6432.

\bibitem{GonzalezRey:1998tk}
F.~Gonzalez-Rey, I.~Y. Park, and K.~Schalm,
\newblock Phys. Lett. {\bf B448}, 37 (1999), arXiv:hep-th/9811155.

\bibitem{Eden:1998hh}
B.~Eden, P.~S. Howe, C.~Schubert, E.~Sokatchev, and P.~C. West,
\newblock Nucl. Phys. {\bf B557}, 355 (1999), arXiv:hep-th/9811172.

\bibitem{Eden:1999kh}
B.~Eden, P.~S. Howe, C.~Schubert, E.~Sokatchev, and P.~C. West,
\newblock Phys. Lett. {\bf B466}, 20 (1999), arXiv:hep-th/9906051.

\bibitem{Eden:2000mv}
B.~Eden, C.~Schubert, and E.~Sokatchev,
\newblock Phys. Lett. {\bf B482}, 309 (2000), arXiv:hep-th/0003096.

\bibitem{Bianchi:2000hn}
M.~Bianchi, S.~Kovacs, G.~Rossi, and Y.~S. Stanev,
\newblock Nucl. Phys. {\bf B584}, 216 (2000), arXiv:hep-th/0003203.

\bibitem{Eden:2011we}
B.~Eden, P.~Heslop, G.~P. Korchemsky, and E.~Sokatchev,
\newblock Nucl. Phys. {\bf B862}, 193 (2012), arXiv:1108.3557.

\bibitem{Eden:2012tu}
B.~Eden, P.~Heslop, G.~P. Korchemsky, and E.~Sokatchev,
\newblock Nucl. Phys. {\bf B862}, 450 (2012), arXiv:1201.5329.

\bibitem{Ambrosio:2013pba}
R.~G. Ambrosio, B.~Eden, T.~Goddard, P.~Heslop, and C.~Taylor,
\newblock JHEP {\bf 01}, 116 (2015), arXiv:1312.1163.

\bibitem{Drummond:2013nda}
J.~Drummond {\em et~al.},
\newblock JHEP {\bf 08}, 133 (2013), arXiv:1303.6909.

\bibitem{Alday:2009zm}
L.~F. Alday, J.~M. Henn, J.~Plefka, and T.~Schuster,
\newblock JHEP {\bf 1001}, 077 (2010), arXiv:0908.0684.

\bibitem{Drummond:2006rz}
J.~Drummond, J.~Henn, V.~Smirnov, and E.~Sokatchev,
\newblock JHEP {\bf 0701}, 064 (2007), arXiv:hep-th/0607160.

\bibitem{Eden:2000bk}
B.~Eden, A.~C. Petkou, C.~Schubert, and E.~Sokatchev,
\newblock Nucl. Phys. {\bf B607}, 191 (2001), arXiv:hep-th/0009106.

\bibitem{CaGe}
G.~Brinkmann, O.~D. Friedrichs, S.~Lisken, A.~Peeters, and N.~Van~Cleemput,
\newblock MATCH Commun. Math. Comput. Chem. {\bf 63}, 533 (2010).

\bibitem{CaronHuot:2012ab}
S.~Caron-Huot and K.~J. Larsen,
\newblock JHEP {\bf 1210}, 026 (2012), arXiv:1205.0801.

\bibitem{Bourjaily:2015jna}
J.~L. Bourjaily and J.~Trnka,
\newblock JHEP {\bf 08}, 119 (2015), arXiv:1505.05886.

\end{thebibliography}

\end{document}